\documentstyle[epsf,axodraw]{aipproc}
\begin{document}
\author{ F. Mamedov}
\address{Department of Physics, State University of New York, Buffalo,
New York 14260} 
\title{  Radiation zeros in weak boson production processes at hadron
colliders} 
\maketitle
\begin{abstract}
{ The Standard Model amplitudes for processes where one or more gauge
bosons are emitted exhibit zeros in the angular distributions. The
theoretical 
and experimental aspects of these radiation amplitude zeros  are 
reviewed and some recent results are discussed. In particular, the zeros 
of the $WZ\gamma$ and $WZZ$ production amplitudes are analyzed. It is 
briefly explained how radiation zeros can be used to test the SM.}
\end{abstract}
\section{INTRODUCTION}
It is well known that 
the Standard Model (SM) amplitudes for many processes exhibit zeros in
the angular distributions. 
Many of these zeros are in the physical region and 
therefore can, in principle, be observed experimentally.  Since the 
zeros leave deep dips in angular distributions for many high energy physics 
processes, they can be used to test the SM. Any
non-standard model physics has a 
strong impact on these dips and will tend to wash them out. 

In Section~2, we discuss the processes
\begin{displaymath}
 pp~(p\bar p) \to W^{\pm}+\gamma+X,
\end{displaymath}
where these zeros were encountered first.
In Section 3 we briefly review classical radiation zeros, which are
a limiting case of the quantum radiation zeros. In Section 4 we 
discuss the conditions which have to be fulfilled in order that 
radiation zeros can occur. In Section 5 we 
consider several examples of radiation zeros. 
Zeros in $WZ$, $WZ\gamma$ and $WZZ$ production at hadron colliders 
and their role in testing the SM are discussed in Section 6.

\section{Radiation zeros in $W\gamma$ Production}
The $pp$ and $p\bar p$ collision processes $pp (p\bar p) \to W^{\pm}+\gamma+X$,
initially proposed as a candidate for measuring the magnetic 
moment of the $W$ boson \cite{W_moment}, exhibit zeros in the angular distribution  at the
parton level \cite{Wgamma_0}.  
\begin{figure}[t]
\begin{tabular}{ll}
\epsfysize=2in
\epsffile[0 560 510 880]{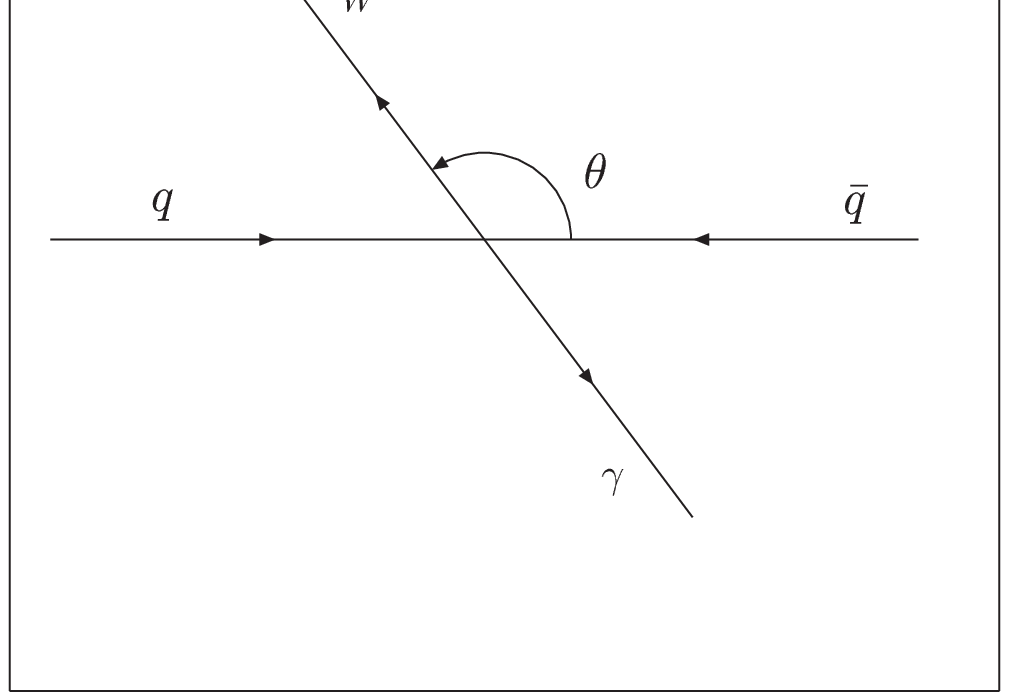} & 
\epsfysize=2in
\epsffile{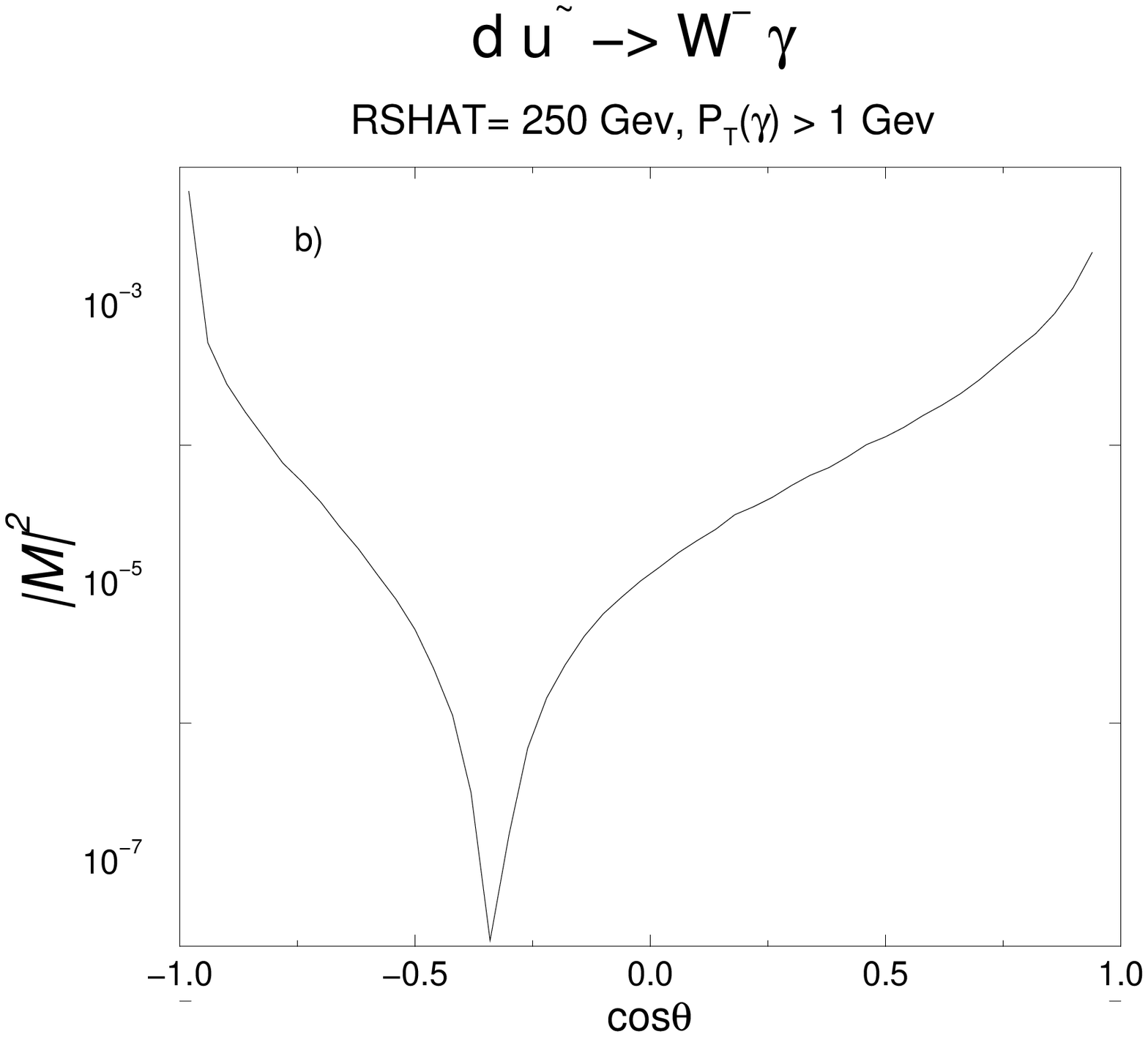} \\
\epsfysize=2in
\end{tabular}
\caption{a) The $q\bar q'\to W\gamma$ collision process and
b) the zero in the angular distribution for $W\gamma$ production. }
\label{parton_Wgamma}
\end{figure}
The position of the zero 
depends only on the charge of the quark (no helicity dependence)~:
\begin{equation}
\cos\theta = -(1+2Q_{i}),
\label{eq:zero}
\end{equation}
where $\theta$ is the angle between $W^{-}$ and $d$ quark
(see Fig. \ref{parton_Wgamma}a),
and $Q_i$ is the electric charge of the quark in units of the proton
charge $e$.

The value of $\cos\theta=-1/3$ (see Fig. \ref{parton_Wgamma}b), obtained from 
Eq. (\ref{eq:zero}), is in fact, characteristic for some other SM based
process amplitudes too, as we will discuss later.  

\begin{figure}[t]
\setlength{\epsfxsize}{2.5 in}
\centerline{\epsffile{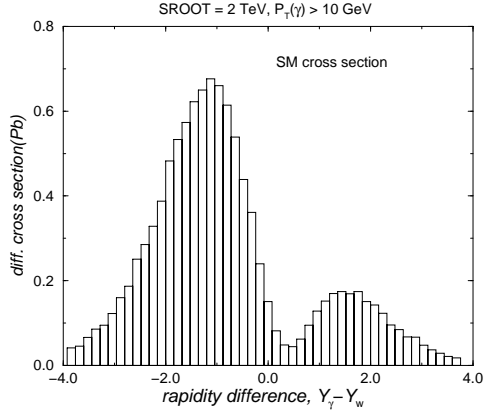}}
\caption{Rapidity difference distribution for the $p\bar p
\to W^{-}\gamma$ process.}
\label{Wgamma}
\end{figure}

From the experimental point of view, the Tevatron collider 
($p\bar p$) is especially well suited to observe the radiation zero
predicted in 
$W\gamma$ production. Sea quark effects tend to wash out the dip caused
by the radiation zero. At Tevatron energies, valence quark effects
dominate and this effect is not a problem and the radiation zero leaves
a clear signature. This is shown in Fig. \ref{Wgamma} where 
we display the distribution of the difference between the rapidities of
the $W$ boson, $Y_W$, and the photon, $Y_\gamma$. The dip at
$Y_{\gamma}-Y_{W}\approx 0.3$ is due to the radiation zero \cite{Baur_1}.
The boost invariant quantity $Y_{\gamma}-Y_{W}$ contains the same
information as the $\cos\theta$ distribution. The $Y_{\gamma}-Y_{W}$
distribution is very similar to the distribution of the rapidity
difference between the photon and the charged lepton originating from
the $W$ decay, which can be readily observed. This is due to the $V-A$
nature of the $W$ fermion antifermion coupling and the fact that $W$'s
in $W\gamma$ production in the SM are strongly polarized: 
the dominant helicity of the $W^{\pm}$ boson in SM $W^{\pm}\gamma$ 
production is $\lambda_{W}=\pm 1$.

\section{Classical Radiation Zeros}
In this section we briefly review the classical radiation zeros, which 
in fact are the limiting case of the so called TYPE~I zeros in quantum
field theory, which will be discussed in the next section \cite{Brown_1}. 

It is well known that the electromagnetic radiation originating from a 
dipole with dipole moment
\begin{equation}
\vec d = \sum_{i}Q_{i}\vec r_{i}
\end{equation}
vanishes, if  
\begin{equation}
Q_{i}/m_{i} = Q_{1}/m_{1},~{\rm for~all}~i. 
\label{eq:Q/m}
\end{equation}

For particles with spin, the magnetic dipole moment is:
\begin{equation}
\vec \mu = \sum_{i}\vec\mu_{i}= \sum_{i}g_{i}\,\frac{Q_{i}}{2m_{i}}\, 
\vec S_{i} .
\label{eq:radzero}
\end{equation} 
If the constants $g_{i}$ are the same for all particles, the condition 
of Eq. (\ref{eq:Q/m}) is 
satisfied, and there are no external torques, 
\begin{equation}
\stackrel{\cdot}{\vec \mu} = 0 
\end{equation}
no radiation occurs.

For the scattering of $relativistic$ particles,
the amplitude for photon radiation in the infrared limit can be written
in the form 
\begin{equation}
A_{IR} = \sum_{i}\frac{Q_{i}}{p_{i} \cdot q}\,\delta_{i}p_{i}\cdot \epsilon
\end{equation}
where $\delta_{i}p_{i}=p_{i}~ (-p_{i})$ before (after) photon
radiation. If
\begin{equation}
\frac{Q_{i}}{p_{i}\cdot q} = {\rm const,~~for~all}~i 
\label{eq:relav}
\end{equation}
where $q$ is the momentum of the photon,
then $A_{IR} = 0$ from momentum conservation and the transversality 
condition $q \cdot \epsilon = 0$. From Eq. (\ref{eq:relav}) 
we observe that the condition for a radiation zero in the case of the
scattering of relativistic particles via the dot product in the
denominator exhibits a dependence on the scattering angle of the photon. 

\section{Radiation Zeros in Quantum Field Theory}
The amplitude zero observed in $W\gamma$ production belongs to 
the family of so-called ``TYPE~I'' zeros  which can be explained as 
a consequence of 
the factorization of the amplitude, shown soon after they were first 
discussed in the literature \cite{Brown_1}-\cite{Goebel}. 

The scattering amplitude for the above mentioned  process can be
obtained starting from the vertices which describe the interaction of
the charged particles, attach a photon to each charged particle 
leg in turn and add all diagrams, as schematically depicted below:
\SetScale{0.8}{
\begin{center}
\begin{picture}(500,120)(-190,-50)
\ArrowLine(-140,50)(-190,50)
\ArrowLine(-225,85)(-190,50)
\ArrowLine(-225,15)(-190,50)
\Text(-100,40)[l]{$\Rightarrow$}
\ArrowLine(-10,50)(-60,50)
\ArrowLine(-95,85)(-60,50)
\ArrowLine(-95,15)(-60,50)
\Photon(-84.75,74.75)(-59.75,74.75){4}{5}
\Vertex(-85,74.75){2}
\Text(10,40)[l]{+}
\ArrowLine(110,50)(60,50)
\ArrowLine(25,85)(60,50)
\ArrowLine(25,15)(60,50)
\Photon(35.25,24.25)(60.25,24.25){4}{5}
\Vertex(34,24.25){2}
\Text(100,40)[l]{+}
\ArrowLine(230,50)(180,50)
\ArrowLine(145,85)(180,50)
\ArrowLine(145,15)(180,50)
\Photon(215,50)(215,75){4}{5}
\Vertex(215,50){2}
\ArrowLine(-190,-20)(-140,-20)
\Text(-100,-17)[l]{$\Rightarrow$}
\Text(-85,-17)[l]{{$charged~ particle~ lines$}}
\Photon(-190,-44)(-165,-44){4}{5}
\Text(-110,-37)[l]{$\Rightarrow$}
\Text(-90,-37)[l]{{$photon~(gauge~ boson)~lines$}}
\end{picture}
\end{center}
}
One can show that  the amplitude (for particles of any 
spin) can be written in the form 
\begin{equation}
M_{\gamma} =  \sum_{i}\frac{A_{i}B_{i}}{C_{i}} 
\end{equation}
where $A_{i}$ and $B_{i}$ are factors which depend on the charge and
polarization. $C_i$ represent the particle propagators. 
This leads to the factorization of the amplitude into 
separately charge dependent and polarization dependent factors:
\begin{equation}
\sum_{i}\frac{A_{i}B_{i}}{C_{i}}=f( A_{i},C_{i})\cdot g( B_{i},C_{i})
\end{equation}
The factorization of the amplitudes holds for any 
\textit{\textbf{gauge theory based vertex}} with no restriction on 
the number of particles, due to the relation between the photon 
(gauge boson) coupling and Poincare invariance. For the complete tree 
level amplitude for a source graph $V_{G}$ consisting of a single vertex 
(no internal lines)
\begin{equation}
M_{\gamma}(V_{G}) = \sum \frac{Q_{i}J_{i}}{p_{i}\cdot q}
\end{equation}
the vertex currents  $J_{i}$ which depend on the polarizations, but not 
on the charges of the particles obey the identity 
\begin{equation}
\sum J_{i}=0
\end{equation}
for all Yang-Mills type vertices, as a result of momentum 
conservation, Lorentz invariance and the Bianchi identity. Thus the 
vertex amplitude $M_{\gamma}(V_{G})$ vanishes if
\begin{equation}
 \frac{Q_{i}}{p_{i}\cdot q}={\rm const,~for~all}~i
\end{equation}
similar to the classical case ($\textit{charge null zone}$). The 
well-known radiation zero occurring in $W\gamma$ production belongs to 
this type.

Since $\sum Q_{i}=0$, the amplitude $M_{\gamma}(V_{G})$ will 
be zero, if 
\begin{equation}
 \frac{J_{i}}{p_{i}\cdot q}={\rm const,~for~all}~i
\end{equation}
or
\begin{equation}
\frac{p_{i} \cdot \epsilon}{p_{i} \cdot q}={\rm const,~for~all}~i
\end{equation}
These zeros correspond to the $\textit{current null zone}$ \cite{Brown_2}. 

From the conditions above we see that the null zones connect 
intrinsic (charge, spin) and space-time properties (Poincare
transformation) of the particles involved. This makes it possible to use 
them in analyzing the structure of the SM.

Radiation zeros can also be explained as a consequence of the 
decoupling theorem~\cite{Brown_3}. The wave function of a
system of particles in an external Yang-Mills field can be 
written as
\begin{equation}
\Psi(x)=ULT\chi(x)
\end{equation} 
where $\chi(x)$ is the free solution of the field equations 
($Q=0$, no gauge boson emission), and 
$ULT$ is the product of the local gauge ($U$), Lorentz ($L$), and 
displacement ($T$) transformations. The null zone condition
\begin{equation}
\prod_{i}(ULT)_{i}=1
\end{equation}
leads to the charge null condition discussed above.

From the condition for the charge null zone we conclude that 
(since $p_{i} \cdot q \geq 0$)
\begin{equation}
\frac{Q_{i}}{Q_{j}} \geq 0, ~{\rm for~all}~i,j.
\end{equation} 
Notice that the zeros will not necessarily be in the physical range of 
the parameters ($-1 \le \cos\theta \le 1$ in the case discussed here).

In $2\to 2$ scattering processes, where one of the final 
particles is a massless, neutral gauge boson, in the relativistic 
limit, the zeros occur at the angle
\begin{equation}
\cos\theta = \frac{Q_{1}-Q_{2}}{Q_{2}+Q_{1}},
\end{equation}  
where $Q_{1}$ and $Q_{2}$ are the charges of the initial state particles.

For the reaction 
$d \bar u \to W^{-} \gamma$
we indeed get $\cos\theta = -\frac{1}{3}$, consistent with the result
from a direct computation of the matrix elements.
The presence of amplitude zeros also requires a gyro-magnetic 
factor of $g=2$. Any anomalous $WW\gamma$ coupling changes the value of 
$g$ and destroys the radiation zero.

\section{Examples of TYPE I and TYPE II Zeros}
Amplitude zeros can be observed in many interesting reactions. In this
section we discuss several examples.

\begin{figure}[t]
\begin{center}
\begin{tabular}{rr}
\epsfysize=2.2in
\epsffile{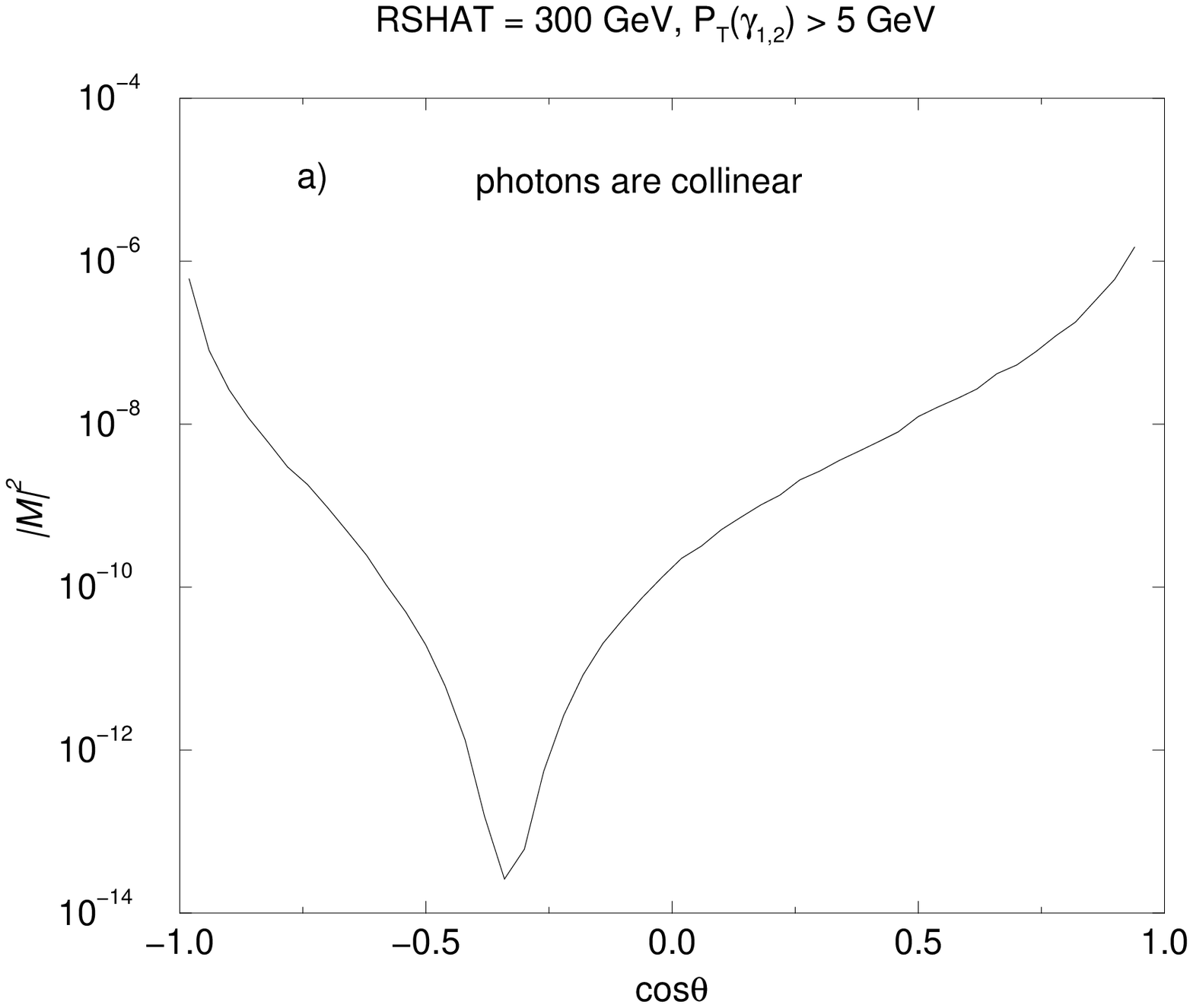} & 
\epsfysize=2in
\epsffile{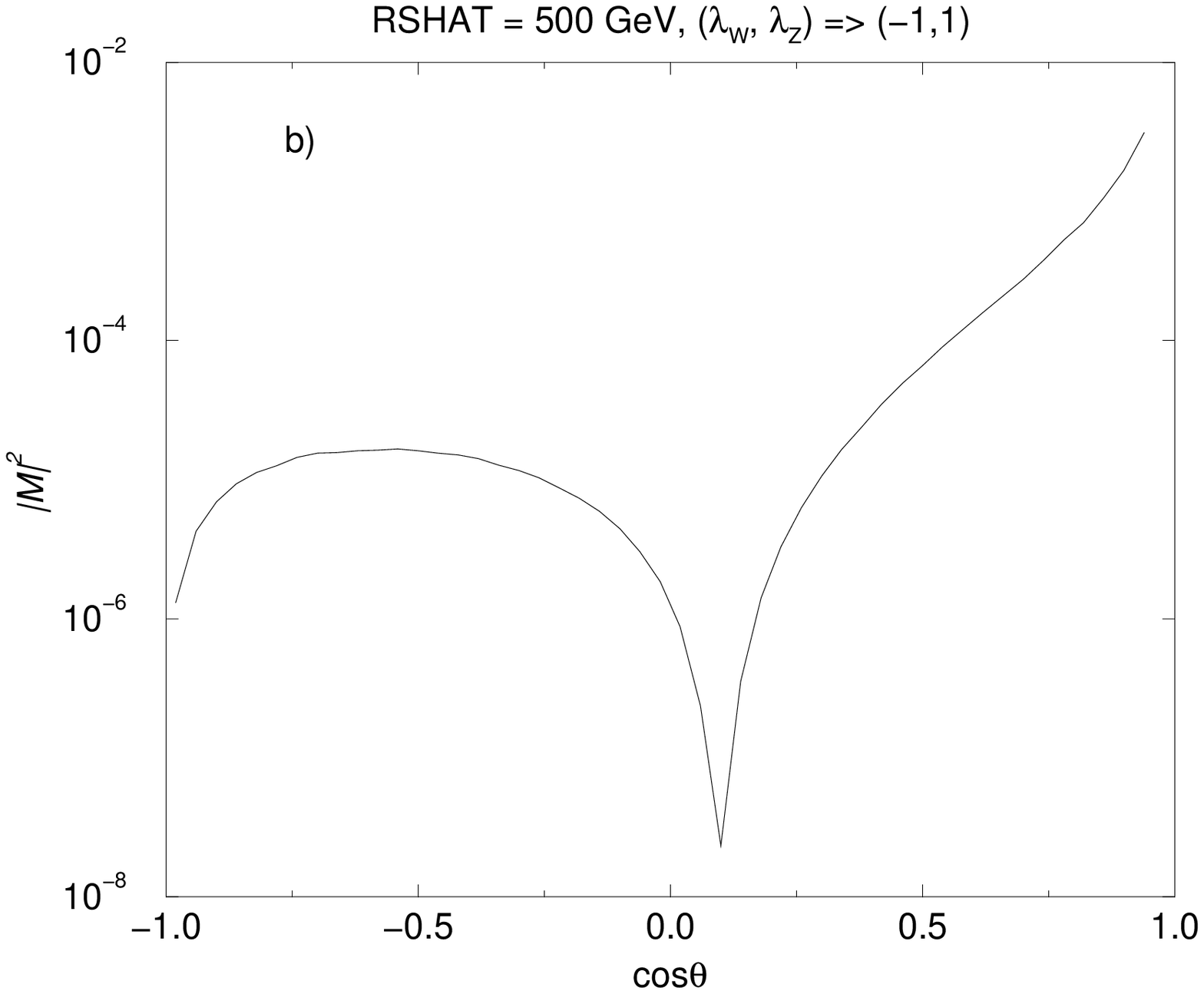} \\
\end{tabular}
\end{center}
\caption{Zeros in a) $W\gamma\gamma$, and  b) $WZ$ production.}
\label{Waa_WZ_zeros}
\end{figure}

A radiation zero exists not only when one massless neutral gauge boson
is emitted, but also if two or more are radiated, provided that the neutral
gauge bosons are all collinear \cite{Brown_4}-\cite{Baur_2}. The radiation zero is located at 
the same scattering angle as  
in the case where only one boson is radiated. This is illustrated in
Fig. \ref{Waa_WZ_zeros}a where we show the squared amplitude as a function of the photon
scattering angle for $W\gamma\gamma$ production at $\sqrt{\hat
s}=300$~GeV. A $p_T(\gamma)> 5$~GeV cut has been imposed to avoid the
infrared singularities associated with photon emission. 

In the case of massive neutral gauge bosons, the 
radiation zero will have an approximate character. Only some of the 
amplitudes will exhibit a true zero, which is weakly energy dependent. 
Fig. \ref{Waa_WZ_zeros}b shows the $d\bar u\to W^-Z$ squared helicity amplitude for 
$\lambda_W=-1$ and $\lambda_Z=+1$ where $\lambda_W$ ($\lambda_Z$) 
is the helicity of the $W$ ($Z$) boson \cite{Baur_3}. 

Recently another type of the zeros (TYPE~II) was discovered 
\cite{Heyssler_1}-\cite{Stirling}
in the physical phase space range for the processes 
\begin{equation}
e^{+}u \to e^{+}u + \gamma,\qquad e^{+}d \to e^{+}d + \gamma,
\end{equation}
and
\begin{equation}
q \bar q \to W^{+}W^{-} \gamma
\end{equation}
Type~II zeros occur only if the emitted photons are located in the 
scattering plane. 

\section{Zeros in $WZ$, $WZ\gamma$ and $WZZ$ production}
\begin{figure}[b]
\begin{tabular}{cr}
\epsfysize=2.3in
\epsffile{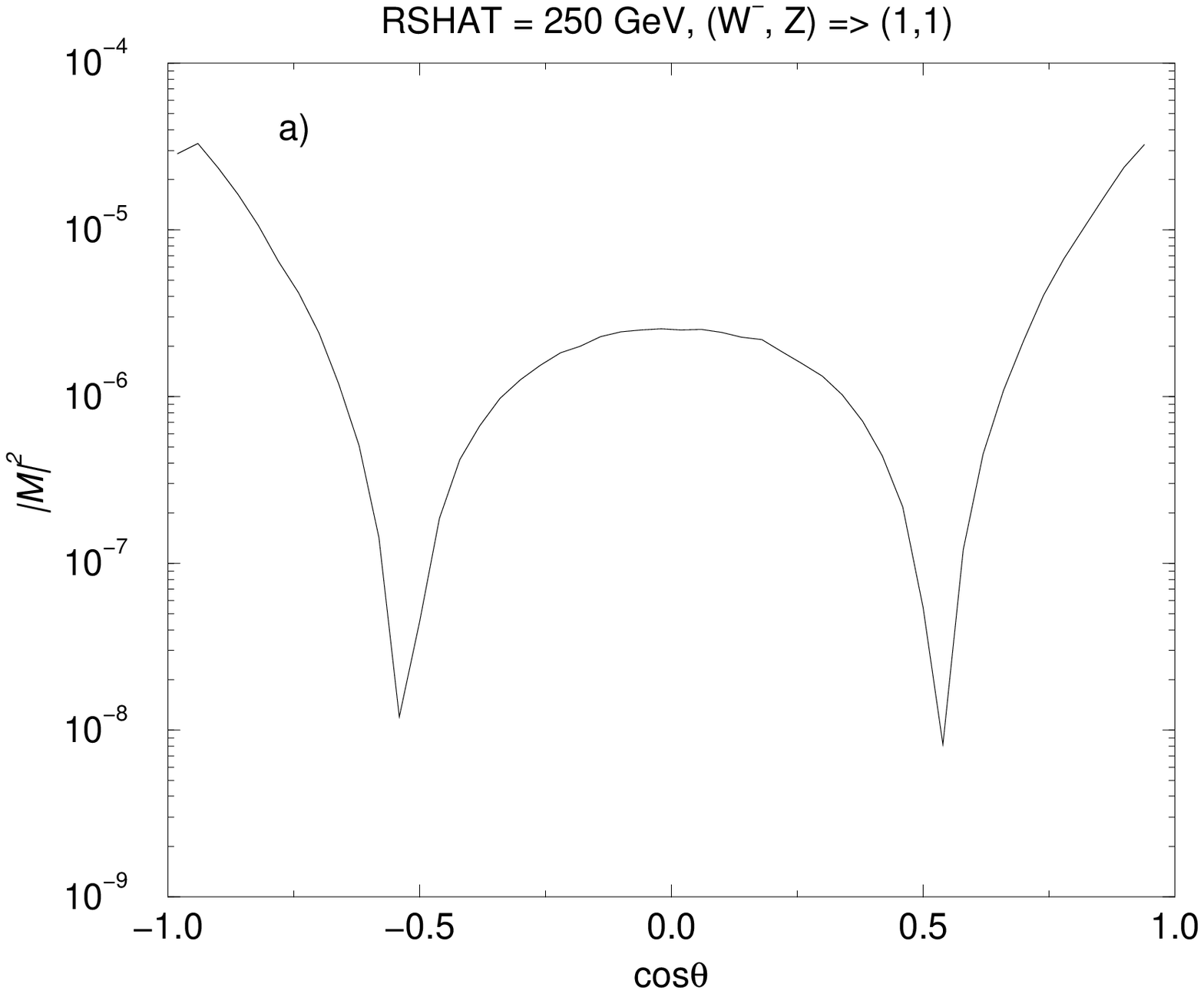} & 
\epsfysize=2.3in
\epsffile{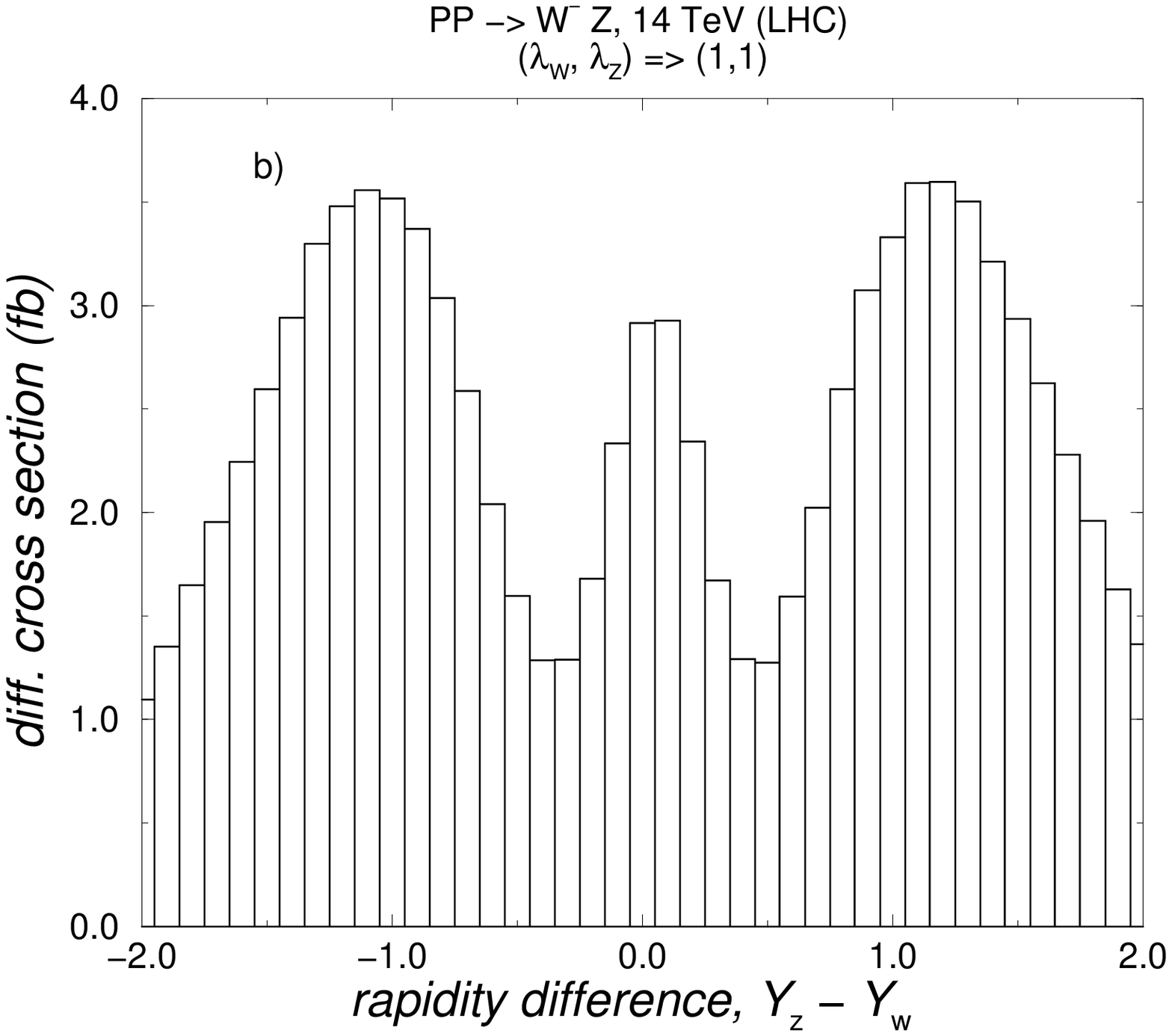} \\
\end{tabular}
\caption{a) The squared $(\lambda_{W},\lambda_{Z})=(1,1)$
amplitude as a function of the scattering angle and b) the $Y_{Z}-Y_{W}$
distribution for $(\lambda_{W},\lambda_{Z})=(1,1)$ in $W^{-}Z$ production
at the LHC.}
\label{WZ_zeros}
\end{figure}

In addition to the approximate radiation zero in $q \bar q \to
W^{\pm}Z$, some of the helicity amplitudes in this process show
additional spin-dependent zeros in the physical region. Some of these are 
approximately symmetric in $\cos\theta$. As an example we show the
$d\bar u\to W^-Z$ squared amplitude for $(\lambda_W,\lambda_Z)=(1,1)$ and
in Fig. \ref{WZ_zeros}a. The $Y_Z-Y_W$ rapidity difference
distribution at the LHC for this helicity combination is shown in Fig.~4b.
Unfortunately the amplitudes for which these
zeros occur contribute negligibly to the cross section so that it will
be very difficult to observe the dips shown in Fig. \ref{WZ_zeros}b.
\\
\begin{figure}[t]
\epsfysize=2.3in
\centerline{\epsffile{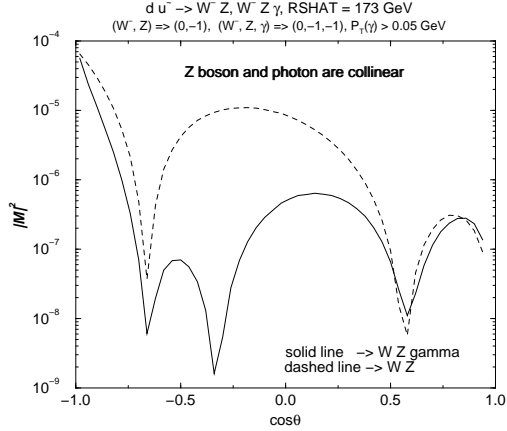}}
\caption{Zeros in $d\bar u\to WZ$ and  $d\bar u\to WZ \gamma$ for
a center of mass energy of $E_{c.m.}=173$~GeV. Shown are the squared 
amplitudes for $(\lambda_{W},\lambda_{Z})=(0,-1)$ and
$(\lambda_{W},\lambda_{Z},\lambda_{\gamma})=(0,-1,-1)$. A 
$p_{T}(\gamma)>0.05$~GeV cut has been imposed to avoid the infrared 
singularity present in
$WZ\gamma$ production. In the $WZ\gamma$ case, the $Z$ boson and the
photon are collinear.}
\label{WZg25}
\end{figure}

The amplitude zeros in the $q \bar q \to WZ\gamma,~WZZ$
processes are especially interesting, as they originate from 
different sources. An example given in Fig. \ref{WZg25} shows that  the 
$WZ\gamma$ production amplitude has an additional zero at
$\cos\theta=-1/3$ besides the zeros present in $WZ$ production. 

In Fig. \ref{WZg25}  we have evaluated the matrix element near the threshold, 
$\sqrt{\hat s}\approx M_{W}+M_{Z}$. At threshold, the 
$W^{-}Z\gamma$ amplitude factorizes into the $W^{-}Z$ amplitude and a
factor describing photon emission. This factor exhibits an energy
independent zero at 
$\cos\theta_{\gamma}= 1/3$ if the $Z$ boson and photon are
collinear. Here, $\theta_{\gamma}$ is 
the angle between the incoming $d$ quark and the photon. 
The other zeros originate from the $WZ$ production amplitude. 

\begin{figure}[tb]
\epsfysize=2.3in
\centerline{\epsffile{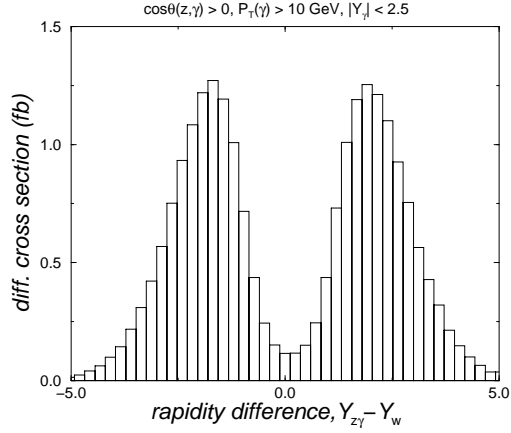}}
\caption{The $Y_{Z\gamma}-Y_{W}$ rapidity difference distribution for
$\cos\theta(Z,\gamma)>0$ in $W^{-}Z\gamma$ production at the LHC.}
\label{WZg01}
\end{figure}

To summarize, the $WZ\gamma$ production amplitudes have a very 
rich structure with zeros originating from three different sources:
\\
1. a radiation zero at $\cos\theta_{\gamma}= 1/3$ connected with
photon radiation
\\
2. the approximate radiation zero occurring in $WZ$ production,
\\
3. the spin dependent zeros in some of the $WZ$ production amplitudes 
discussed above.
\\
The zero present at $\cos\theta_{\gamma}= 1/3$ leads to a clear dip in
the $Y_{Z\gamma}-Y_{W}$ rapidity difference distribution ($Y_{Z\gamma}$ is 
the rapidity of the $Z\gamma$ system) if the cosine of the angle between 
the $Z$ boson and photon is restricted to $\cos\theta(Z,\gamma)>0$. The 
$Y_{Z\gamma}-Y_{W}$ distribution at the LHC is shown in Fig. \ref{WZg01}.
\begin{figure}[bt]
\begin{center}
\begin{tabular}{cr}
\epsfysize=2.2in
\epsffile{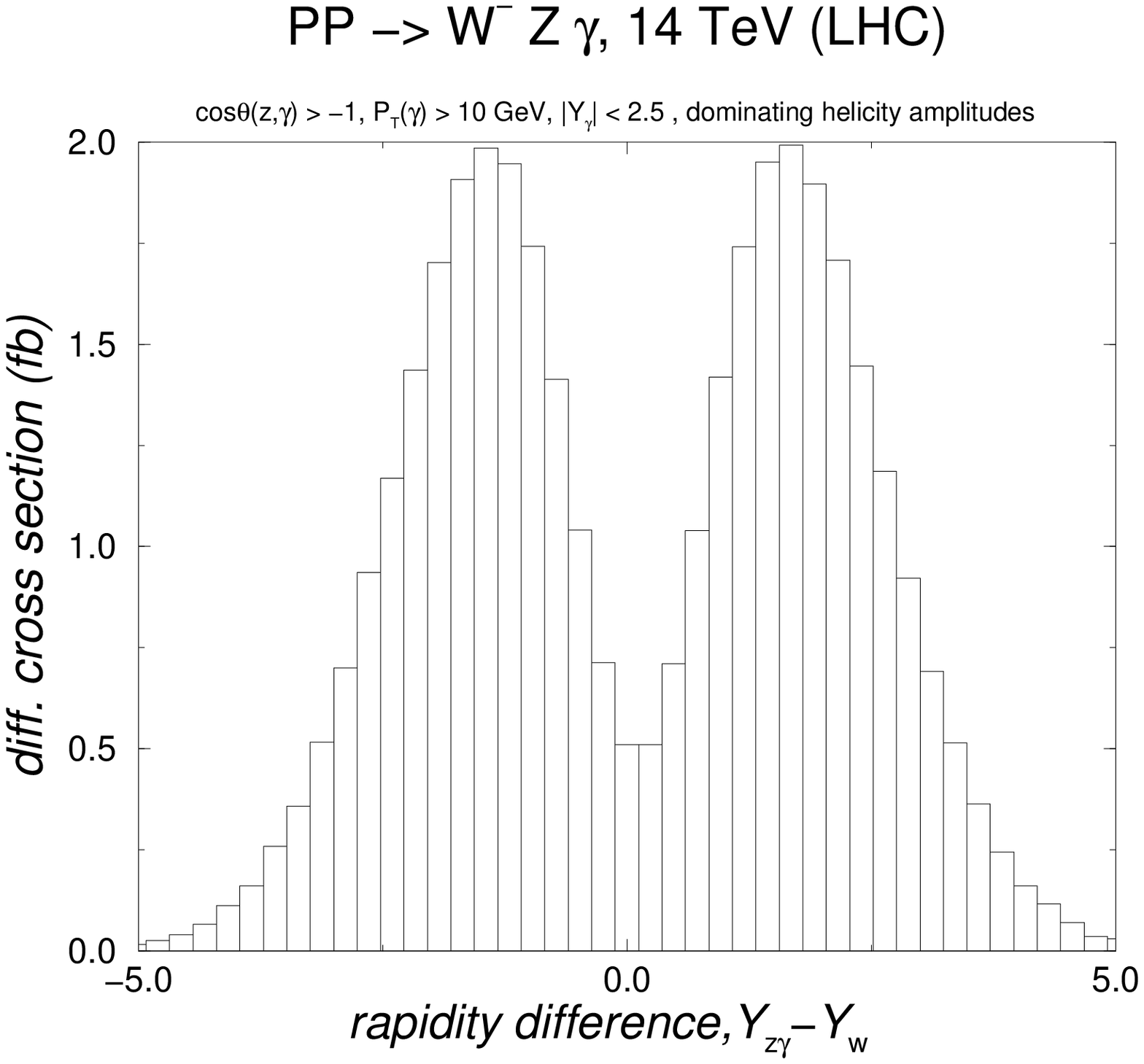} & 
\epsfysize=2.2in
\epsffile{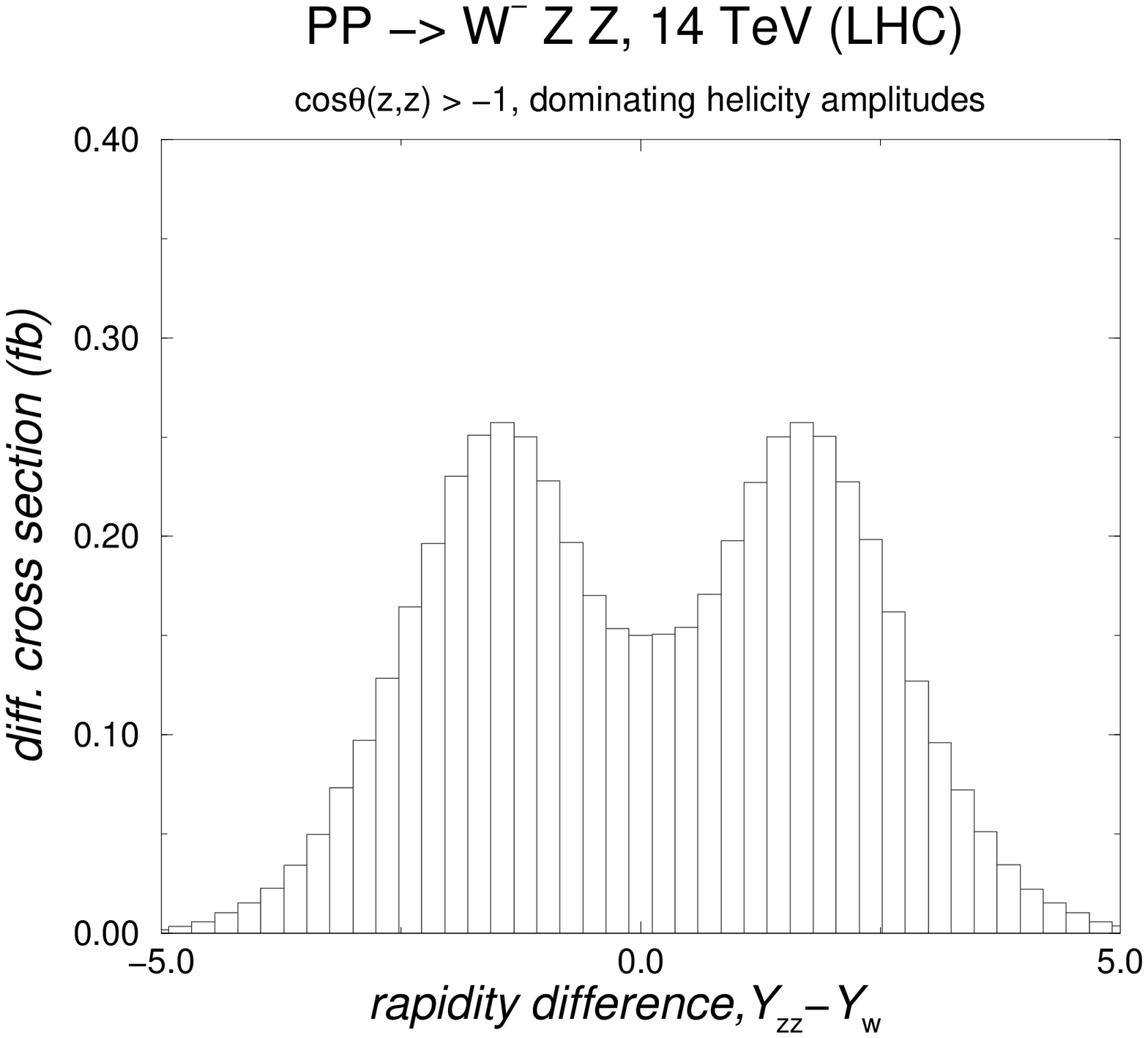} \\
\end{tabular}
\end{center}
\caption{Comparison of the $Y_{Z\gamma}-Y_{W}$ distribution in
$WZ\gamma$ production, and the $Y_{ZZ}-Y_{W}$ distribution in $WZZ$
production at the LHC.}
\label{wzgh_wzzh}
\end{figure}

Similar to $WZ$ production, 
the amplitudes for $WZZ$ production exhibit an approximate radiation
zero if the two $Z$ bosons are collinear. It leads to a dip in the 
$Y_{ZZ}-Y_{W}$ distribution, which, however, is much less pronounced than
the corresponding dip in the $Y_{Z\gamma}-Y_{W}$ distribution in
$WZ\gamma$ production. A comparison of the two distributions at the LHC 
is shown in Fig. \ref{wzgh_wzzh}. Only the helicity amplitudes which give the
dominating contributions to the cross section have been taken into
account here.

\begin{figure}[tb]
\epsfysize=2.3in
\centerline{\epsffile{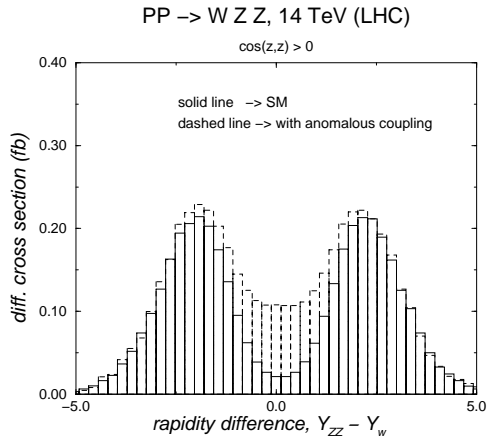}}
\caption{The $Y_{ZZ}-Y_{W}$ distribution in $WZZ$ 
production at the LHC. The solid histogram shows the SM prediction. The
dashed histogram shows the $Y_{ZZ}-Y_{W}$ distribution for $g_c=0.05$.}
\label{wzg_a}
\end{figure}

The radiation zeros can be used for testing the models beyond
the SM \cite{Baur_4}-\cite{Gangemi}. Any anomalous coupling term added to the Lagrangian
causes the dips originating from the amplitude zeros to be washed out. 
In Fig. \ref{wzg_a} we show\footnote{Form-factors \cite{Baur_4} were not used for this graph of 
the illustrative purpose} how an anomalous coupling described by the effective
Lagrangian 
\begin{equation}
 L~=~\frac{g_{c}g^{2}}{cos^{2}\theta_{W}} W_{\mu}Z^{\mu}W_{\nu}Z^{\nu}
\end{equation}
with $g_{c}=0.05$ affects the $Y_{ZZ}-Y_{W}$ distribution in $WZZ$ 
production \cite{Belanger}.

\section{CONCLUSION}
Radiation zeros are a consequence of the factorization of  
tree level amplitudes in the SM. The zeros occur for processes where 
one or more neutral, massless gauge bosons are radiated. If the emitted 
gauge boson is massive, only some of the helicity amplitudes exhibit zeros.
Radiation zeros leave deep measurable 
dips in rapidity difference distributions for many hadron collider
processes, such as $W\gamma$, $WZ$, 
$WZZ$ and $WZ\gamma$ production. 
Since anomalous coupling terms destroy the radiation zeros, they
can be used for probing the SM.

\section*{ACKNOWLEDGMENTS}
The author had several helpful discussions  with Ulrich Baur 
and Anja Werthenbach on the topic of this work.

\end{document}